\documentclass[12pt]{article}

\usepackage{epsfig}
\usepackage{amssymb,amsmath}
\usepackage{setspace}

\newcommand{\bea}{\begin{eqnarray}}
\newcommand{\eea}{\end{eqnarray}}
\def\tr{\mathrm{tr}}
 \makeatletter
\def\alt{\mathrel{\mathpalette\gl@align<}}
\def\agt{\mathrel{\mathpalette\gl@align>}}
\def\gl@align#1#2{\lower.6ex\vbox{\baselineskip\z@skip\lineskip\z@
\ialign{$\m@th#1\hfil##\hfil$\crcr#2\crcr\sim\crcr}}} \makeatother

\begin{document}
%
\vspace*{1.0cm}

\begin{center}
\baselineskip 20pt {\Large\bf
125 GeV Higgs, Type III Seesaw and Gauge-Higgs Unification 
}
\vspace{1cm}

{\large Bin He$^{a}$\footnote{ E-mail:hebin@udel.edu}, 
Nobuchika Okada$^{b}$\footnote{E-mail:okadan@ua.edu}
and Qaisar Shafi$^{a}$\footnote{ E-mail:shafi@bartol.udel.edu}
} \vspace{.5cm}

{\baselineskip 20pt \it
$^a$Bartol Research Institute, Department of Physics and Astronomy, \\
University of Delaware, Newark, DE 19716, USA \\
\vspace{2mm} 
$^b$Department of Physics and Astronomy, 
University of Alabama, Tuscaloosa, AL 35487, USA 
}
\vspace{.5cm}

\vspace{1.5cm} {\bf Abstract}
\end{center}

Recently, both the ATLAS and CMS experiments 
 have observed an excess of events 
 that could be the first evidence for a 125 GeV Higgs boson. This is a few GeV 
 below the (absolute) vacuum stability bound on the Higgs mass
 in the Standard Model (SM), assuming a Planck mass ultraviolet (UV) cutoff.
In this paper, we study some implications of a  125 GeV Higgs boson for new physics in terms of the vacuum stability bound. 
We first consider the seesaw extension of the SM and 
 find that in type III seesaw, the vacuum stability 
 bound on the Higgs mass can be as low as 125 GeV 
 for the seesaw scale around a TeV. 
Next we dicuss some alternative new physics models
 which provide an effective ultraviolet cutoff lower than the Planck mass. 
An effective cutoff $\Lambda \simeq 10^{11}$ GeV 
 leads to a vacuum stability bound on the Higgs mass of 125 GeV. 
In a gauge-Higgs unification scenario 
 with five-dimensional flat spacetime, 
 the so-called gauge-Higgs condition allows us to predict 
 a Higgs mass of 125 GeV, with the compactification scale 
 of the extra-dimension being identified as 
 the cutoff scale $\Lambda \simeq 10^{11}$ GeV. 
Identifying the compactification scale with
 the unification scale of the SM SU(2) gauge 
coupling and the top quark Yukawa coupling yields a Higgs mass of $121\pm 2$ GeV.

\thispagestyle{empty}

\newpage

\addtocounter{page}{-1}

\baselineskip 18pt

\section{Introduction}

The discovery of the Higgs boson is a major goal of 
 the physics program at the Large Hadron Collider (LHC), 
 in order to confirm the origin of the electroweak 
 symmetry breaking and the mechanism of particle mass generation. 
The ATLAS \cite{ATLAS} and CMS \cite{CMS} experiments 
 have reported an excess of events that could be the first evidence 
 of the Higgs boson with a mass of around 125 GeV \cite{Moriond}. 
Recent analysis by the Tevatron experiments \cite{Moriond} 
 supports the above observations.

A value of 125 GeV is quite interesting from the viewpoint of the vacuum stability bound on the Higgs boson mass \cite{stability1}. 
In the SM, the Higgs boson mass is determined by the self-coupling 
 of the Higgs doublet, so that we can analyze the high energy 
 behavior of the self-coupling by using the renormalization 
 group equations (RGEs). 
For a relatively light Higgs boson, the self-coupling becomes 
 negative in its RGE running at some high energy, 
 which implies an instability of the effective Higgs potential. 
If we require that the running self-coupling remains positive 
 below a given cutoff scale, we obtain a lower bound on 
 the Higgs boson mass, known as the (absolute) vacuum stability bound.
 
It would seem natural to adopt the reduced Planck mass as the cutoff scale, 
 in which case the vacuum stability bound is found to be close to
 129 GeV for a top quark mass of 173.2 GeV \cite{Mt}. Taking into account the uncertainty on the top quark mass ($M_t=173.2\pm0.9$ GeV),
 
\begin{equation}
m_H\geq 128.9~{\rm GeV} + 1.9~{\rm GeV} \left( \frac{M_t-173.2 ~{\rm GeV}}{0.9~{\rm GeV}} \right). 
\end{equation}

If the observed excess of events around 125 GeV actually is 
evidence for the Higgs boson, we may entertain two possibilities 
 for lowering the vacuum stability bound of 129 GeV.
One possibility is that the RGE running of the quartic self-coupling 
 is altered from the one in the SM, keeping the reduced Planck mass cutoff. This means that new particles are involved in the RGEs 
 at energies below the reduced Planck mass. 
Another possibility is that the effective cutoff scale lies suitably below the reduced Planck scale, while all the RGEs of the SM 
remain unaltered.
In general, one also could consider a combination of 
 these two possibilities. 
In any case, new physics beyond the SM should play  
 a crucial role to reconcile the discrepancy 
 between 125 GeV and 129 GeV.

In this paper, we study the implications of a 125 GeV Higgs boson for new physics from the viewpoint of the vacuum stability bound. 
We first consider a seesaw extension of the SM 
 where the RGE running of the self coupling is altered 
 by the presence of new particles. 
We will see that the type of seesaw as well as the seesaw scale 
 are restricted in order to realize a Higgs mass of 125 GeV, 
 with the reduced Planck scale cutoff. 
For a different possibility, we will consider physics models which can provide an effective cutoff 
 scale lower than the reduced Planck mass. 
As a very interesting example, we investigate gauge-Higgs 
 unification in flat five-dimensional (5D) spacetime. 
In this model, the effective cutoff scale is identified as 
 the compactification scale of the fifth dimension 
 and a Higgs mass of 125 GeV determines the compactification scale. 
We find a Higgs mass prediction close to 125 GeV if the compactification scale is identified with the unification 
 scale of the top quark Yukawa and SU(2) gauge couplings.

\section{Seesaw Extended Standard Model}

The seesaw mechanism is a simple and promising extension 
 of the SM to incorporate the neutrino masses and flavor 
 mixings observed in solar and atmospheric neutrino oscillations. 
There are three main seesaw extensions of the SM, 
 type I \cite{seesawI}, type II \cite{seesawII}, and 
 type III \cite{seesawIII}, in which singlet right-handed neutrinos, SU(2) triplet scalar, 
 and SU(2) triplet right-handed neutrinos, respectively, are introduced. 
These new particles contribute to the RGEs at energies 
 higher than the seesaw scale and as a result, 
 the vacuum stability bound can be significantly altered. 
Some time ago, the important implications of the various seesaw models 
 (type I \cite{HMass-typeI, HMass-typeIII}, 
 type II \cite{HMass-typeII} and type III \cite{HMass-typeIII}) on the Higgs boson mass have been investigated 
 with the Planck mass cutoff.
In these papers, in addition to the vacuum stability bound, 
 the perturbativity bound, given by the condition 
 that the Higgs self-coupling remains perturbative below the Planck scale, has also been investigated. 
For both type I and III, it has been shown that 
 the window for the Higgs boson mass between the vacuum stability 
 and the perturbativity bounds becomes narrower and is eventually 
 closed by the dramatic rise of the vacuum stability bound, as the neutrino Dirac Yukawa coupling becomes larger. 
  
For lower values of the seesaw scale, the neutrino Dirac Yukawa coupling is small, and there is little effect from this coupling\footnote{
In a general parameterization, the neutrino Dirac
Yukawa coupling can be large and affect the Higgs mass
bounds~\cite{Rodejohann:2012px},
although fine-tuning of parameters is required
to realize the neutrino oscillation data.
}.
However, there is a remarkable difference between type I and type III because the right-handed neutrinos in type III are SU(2) triplets. 
As shown in \cite{HMass-typeIII}, the vacuum stability bound becomes lower for decreasing seesaw scale. 
This is because the SU(2) triplet neutrinos change
 the RGE running of the SU(2) gauge coupling. 
In type II seesaw, the perturbativity bound receives 
 a drastic reduction due to interactions 
 between the Higgs doublet and the SU(2) triplet scalar. 
As a result, the window for the Higgs boson mass 
 between the vacuum stability and the perturbativity bounds 
 becomes narrower and is eventually closed 
 by the dramatic fall of the perturbativity bound 
 for larger values of the scalar couplings. 
The vacuum stability bound also receives a dramatic reduction 
 when the seesaw scale is low. 
It has been shown in \cite{HMass-typeII} that 
 in type II seesaw, the Higgs stability bound 
 becomes even lower than the LEP2 Higgs mass bound 
 of 114.4 GeV \cite{LEP2} for a seesaw scale of around 1 TeV.

In the light of the recent LHC results suggesting a Higgs mass 
close to 125 GeV, type II and III seesaw models are interesting 
 because in both cases, the vacuum stability bound 
 can be lower than the SM prediction of 129 GeV. 
Since type II seesaw involves many free parameters, 
 there is a wide range of parameter regions 
 which yield a vacuum stability bound of 125 GeV. 
For this reason, in this paper we consider type III seesaw in detail. 
In low scale type III seesaw compatible with a Higgs mass of 125 GeV, the neutrino Dirac Yukawa coupling is too small 
 to play a role in the RGE running of the Higgs self-coupling. 
Therefore, the only free parameters involved in our analysis 
 are the
  masses of the SU(2) triplet right-handed neutrinos. 
We analyze three cases with 1, 2 and 3 generations 
 of the triplet neutrinos. 
Although at least 2 right-handed neutrinos are necessary 
 to reproduce the neutrino oscillation data, 
 we also analyze the 1 generation case for completeness\footnote{
 One may consider a combination of type I and type III 
 to reproduce the neutrino oscillation data. 
 Since a light singlet neutrino  has no effect on RGEs, 
 our result with one triplet neutrino corresponds to this case. 
 }. 
For simplicity, we consider a degenerate mass spectrum 
 for the triplet neutrinos. 
This assumption is reasonable if we consider thermal 
 leptogenesis \cite{LG}, 
 where the CP-violating out-of-equilibrium decay of 
 the right-handed neutrinos generates the lepton asymmetry in the universe. 
It is known that in order to generate a sufficient amount of baryon asymmetry, the seesaw scale should be higher than 
 $10^{10}$ GeV \cite{LGbound}, otherwise a certain enhancement 
 mechanism of the CP-asymmetry parameter is necessary. 
As we will see in the following, the seesaw scale turns out 
 to be much lower than the above bound and hence, 
 the so-called resonant leptogenesis \cite{RLG} is relevant 
 to our case, where the CP-asymmetry parameter is enhanced 
 by right-handed neutrinos that are almost degenerate in mass.

Let us now analyze the vacuum stability bound in type III seesaw extended SM. 
We introduce $N$ generations of mass degenerate right-handed neutrinos 
 which transform as $({\bf 3}, 0)$ under the electroweak gauge group 
 SU(2)$\times$U(1)$_Y$: 
\bea
 \psi_i = \sum_a \frac{\sigma^a}{2} \psi_i^a
=\frac{1}{2}
 \left( \begin{array}{cc}
    \psi^0_i & \sqrt{2} \psi^{+}_i   \\
    \sqrt{2} \psi^{-}_i & - \psi^0_i   \\
 \end{array}\right) .
\eea 
The terms in the Lagrangian relevant for  the seesaw mechanism 
 are given by 
\bea 
 {\cal L} \supset 
 - y_{ij} \overline{\ell_i} \psi_j \Phi 
 - M_R \tr \left[ \overline{\psi^c_i} \psi_i \right],   
 \label{Yukawa}  
\eea
 where $\ell_i$ is the $i$-th generation SM lepton doublet ($i=1,2,3$), 
 $\Phi$ is the SM Higgs doublet with a U(1)$_Y$ charge $-1/2$, 
 and $M_R$ is the common mass for the triplet neutrinos. 
The light neutrino mass matrix obtained via type III seesaw 
mechanism is given by
\bea
  {\bf M}_\nu = \frac{v^2}{8 M_R} {\bf Y} {\bf Y}^T, 
\eea 
where $v=246$ GeV is the vacuum expectation value of the Higgs doublet, 
and ${\bf Y} = y_{ij}$ is a 3$\times N$ Yukawa matrix. 
It is natural to expect the light neutrino mass scale to be 
 ${\cal O}(\sqrt{\Delta m_{23}}) \sim0.05$ eV, 
 where $\Delta m_{23} =2.43 \times 10^{-3}$ eV$^2$ \cite{PDG} 
 is given by the atmospheric neutrino oscillation data. 
Using the seesaw formula, we find $y_{ij} \ll 1 $ 
 for $M_R \ll 10^{15}$ GeV. 
This is the case we analyze here, and so the neutrino Dirac Yukawa 
 coupling has essentially no effect on our results. 
In the following analysis, we employ RGEs at two-loop level.

For a renormalization scale $\mu < M_R$,
 the heavy neutrinos are decoupled, 
 and there is no effect on the RGEs for the SM couplings. 
For the three SM gauge couplings $g_i$ ($i=1,2,3$), we have 
\bea
 \frac{d g_i}{d \ln \mu} =
 \frac{b_i}{16 \pi^2} g_i^3 +\frac{g_i^3}{(16\pi^2)^2}
  \left( \sum_{j=1}^3 B_{ij}g_j^2 - C_i y_t^2   \right),
\label{gauge}
\eea
 where
\bea
b_i = \left(\frac{41}{10},-\frac{19}{6},-7\right),~~~~
 { B_{ij}} =
 \left(
  \begin{array}{ccc}
  \frac{199}{50}& \frac{27}{10}&\frac{44}{5}\\
 \frac{9}{10} & \frac{35}{6}&12 \\
 \frac{11}{10}&\frac{9}{2}&-26
  \end{array}
 \right),  ~~~~
C_i=\left( \frac{17}{10}, \frac{3}{2}, 2 \right),
\label{beta} 
\eea 
and we have included the contribution 
 from the top Yukawa coupling ($y_t$). 
We use the top quark pole mass $M_t=173.2$ GeV \cite{Mt}
 and the strong coupling constant at the Z-pole ($M_Z$) 
 $\alpha_S=0.1193$ \cite{Gfitter}. 
For the top Yukawa coupling, we have 
\bea \label{ty}
 \frac{d y_t}{d \ln \mu}
 = y_t  \left(
 \frac{1}{16 \pi^2} \beta_t^{(1)} + \frac{1}{(16 \pi^2)^2} \beta_t^{(2)}
 \right).
\eea
Here the one-loop contribution is
\bea
 \beta_t^{(1)} =  \frac{9}{2} y_t^2 -
  \left(
    \frac{17}{20} g_1^2 + \frac{9}{4} g_2^2 + 8 g_3^2
  \right) ,
\label{topYukawa-1}
\eea
while the two-loop contribution is given by \cite{RGE} 
\bea
\beta_t^{(2)} &=&
 -12 y_t^4 +   \left(
    \frac{393}{80} g_1^2 + \frac{225}{16} g_2^2  + 36 g_3^2
   \right)  y_t^2  \nonumber \\
 &&+ \frac{1187}{600} g_1^4 - \frac{9}{20} g_1^2 g_2^2 +
  \frac{19}{15} g_1^2 g_3^2
  - \frac{23}{4}  g_2^4  + 9  g_2^2 g_3^2  - 108 g_3^4 \nonumber \\
 &&+ \frac{3}{2} \lambda^2 - 6 \lambda y_t^2 .
\label{topYukawa-2}
\eea

In solving the RGE for the top Yukawa coupling, 
 its value at $\mu=M_t$ is determined from the relation
 between the pole mass and the running Yukawa coupling
 \cite{Pole-MSbar, Pole-MSbar2},
\bea
  M_t \simeq m_t(M_t)
 \left( 1 + \frac{4}{3} \frac{\alpha_3(M_t)}{\pi}
          + 11  \left( \frac{\alpha_3(M_t)}{\pi} \right)^2
          - \left( \frac{m_t(M_t)}{2 \pi v}  \right)^2
 \right),
\eea
 with $ y_t(M_t) = \sqrt{2} m_t(M_t)/v$. 
Here, the second and third terms in parentheses correspond to
 one- and two-loop QCD corrections, respectively,
 while the fourth term comes from the electroweak corrections 
 at one-loop level.

The RGE for the Higgs self-coupling is given by \cite{RGE},
\bea
\frac{d \lambda}{d \ln \mu}
 =   \frac{1}{16 \pi^2} \beta_\lambda^{(1)}
   + \frac{1}{(16 \pi^2)^2}  \beta_\lambda^{(2)},
\label{self}
\eea
with
\bea
 \beta_\lambda^{(1)} &=& 12 \lambda^2 -
 \left(  \frac{9}{5} g_1^2+ 9 g_2^2  \right) \lambda
 + \frac{9}{4}  \left(
 \frac{3}{25} g_1^4 + \frac{2}{5} g_1^2 g_2^2 +g_2^4
 \right) + 12 y_t^2 \lambda  - 12 y_t^4 ,
\label{self-1}
\eea
and
\bea
  \beta_\lambda^{(2)} &=&
 -78 \lambda^3  + 18 \left( \frac{3}{5} g_1^2 + 3 g_2^2 \right) \lambda^2
 - \left( \frac{73}{8} g_2^4  - \frac{117}{20} g_1^2 g_2^2
 - \frac{1887}{200} g_1^4  \right) \lambda - 3 \lambda y_t^4
 \nonumber \\
 &&+ \frac{305}{8} g_2^6 - \frac{289}{40} g_1^2 g_2^4
 - \frac{1677}{200} g_1^4 g_2^2 - \frac{3411}{1000} g_1^6
 - 64 g_3^2 y_t^4 - \frac{16}{5} g_1^2 y_t^4
 - \frac{9}{2} g_2^4 y_t^2
 \nonumber \\
 && + 10 \lambda \left(
  \frac{17}{20} g_1^2 + \frac{9}{4} g_2^2 + 8 g_3^2 \right) y_t^2
 -\frac{3}{5} g_1^2 \left(\frac{57}{10} g_1^2 - 21 g_2^2 \right)
  y_t^2  - 72 \lambda^2 y_t^2  + 60 y_t^6.
\label{self-2}
\eea

The Higgs boson pole mass $m_H$ is determined 
 through a one-loop effective potential improved by two-loop RGEs. 
The second derivative of the effective potential 
 at the potential minimum leads to \cite{HiggsPole} 
\bea
  m_H^2 &=& \lambda \zeta^2 v^2 + \frac{3}{64 \pi ^2} \zeta^2 v^2
 \left\lbrace  g^4_2
 \left( \log{\frac{g^2_2 \zeta^2 v^2}{4 \mu^2}} + \frac{2}{3}\right) \right.  
 \nonumber \\
 &+& \frac{1}{2}\left(g^2_2+ \frac{3}{5}g^2_1 \right)^2 
 \left[\log{ \frac{ \left(g^2_2+ \frac{3}{5}g^2_1 \right)
 \zeta^2 v^2 }{4 \mu^2}}+
 \frac{2}{3}\right] - \left. 8y^4_t \log{\frac{y^2_t \zeta^2v^2}{2\mu^2}}
 \right\rbrace ,  
\label{HiggsPole}
\eea
 where $\zeta =\exp{\left( -\int_{M_Z}^\mu 
 \frac{\gamma(\mu)}{\mu} d\mu\right)}$,  
 with the anomalous dimension $\gamma$ of the Higgs doublet 
 evaluated at two-loop level. 
All running parameters are evaluated at $\mu=m_H$, 
 and the Higgs boson mass is determined as the root 
 of this equation. 
We have checked that our results on 
 the Higgs boson mass bounds for the SM case coincide 
 with the ones obtained in recent analysis \cite{Ellis:2009tp}.

For the renormalization scale $\mu \geq M_R$, 
 the SM RGEs should be modified to include contributions
 from the triplet neutrinos in type III seesaw,
 so that the effectively RGE evolution of the Higgs self-coupling is altered. 
For simplicity, we consider only one-loop corrections 
 from the heavy neutrinos. 
As we have discussed above, the neutrino Dirac Yukawa coupling 
 is very small and its effect in our analysis is negligible. 
Therefore, the presence of the triplet neutrinos  
 only modifies the SU(2) gauge coupling 
 beta function: 
\bea 
 b_2 =- \frac{19}{6} \to - \frac{19}{6} + \frac{4}{3} N,  
\eea  
 corresponding to N generations of triplet neutrinos.

\begin{figure}[ht]
\begin{center}
{\includegraphics*[width=.7\linewidth]{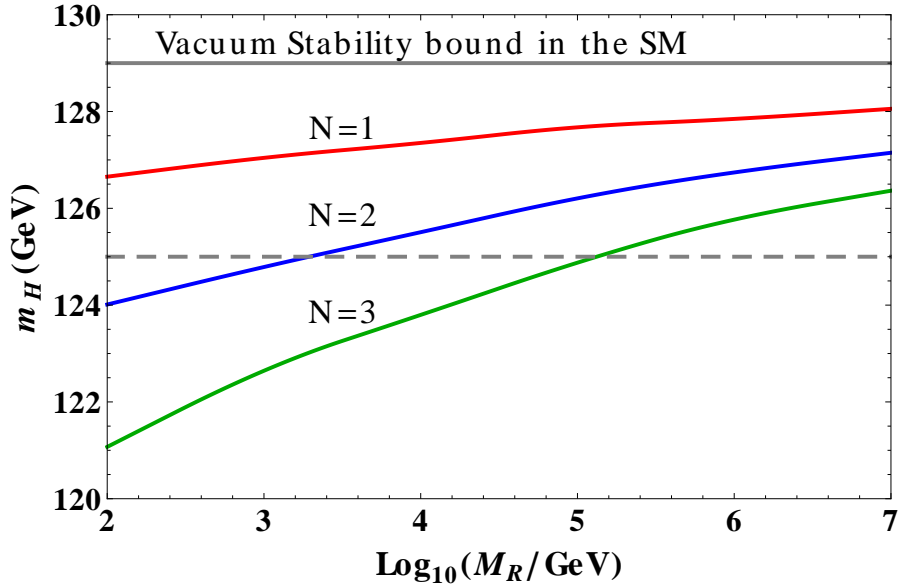}}
\caption{
The vacuum stability bound on the Higgs boson mass 
 as a function of the triplet neutrino  mass 
 for $N=1,2$ and $3$ generations, with reduced 
 Planck mass cutoff. 
We have taken $M_t=173.2$ GeV. 
The horizontal solid line denotes the vacuum stability 
 bound in the SM while $m_H=125$ GeV is shown as the dashed-line. 
}
\end{center}
\end{figure}

\begin{table}[t]
\begin{center}
\begin{tabular}{|c|c|ccc|}
\hline
$N=2$ & $M_t (\rm{GeV})$ & $172.3$  & $173.2$  & $174.1$   \\
      & $M_R (\rm{GeV})$ 
  & $3.2 \times 10^6$ & $1.6 \times 10^3$ & $-$   \\ 
\hline
\hline
$N=3$ & $M_t (\rm{GeV})$ & $172.3$  & $173.2$  & $174.1$   \\
      & $M_R (\rm{GeV})$ 
  & $1.0 \times 10^8$ & $1.6 \times 10^5$ & $3.2 \times 10^3$ \\ 
\hline
\end{tabular}
\end{center}
\caption{
The seesaw scales which give the vacuum stability bound 
 $m_H=125$ GeV for varying $M_t$ values and the reduced Planck mass cutoff. 
}
\end{table}

In Fig.~1, we show the vacuum stability bound on the Higgs boson mass 
 as a function of the triplet neutrino  mass 
 for $N=1,2$ and $3$, respectively. 
Here we have used $M_t=173.2$ GeV and the cutoff scale 
 is taken to be the reduced Planck mass $M_P=2.44 \times 10^{18}$ GeV.
We see that the presence of triplet neutrinos 
 lowers the the resultant Higgs mass from 
 the SM value $\simeq 129$ GeV. 
Note that at least two generations of triplet neutrinos 
 are necessary to yield a Higgs mass of 125 GeV. 
Interestingly, two generation is also the minimal number required to reproduce the neutrino oscillation data.

For $N=2$ and $3$, respectively, we list in Table~1 the values of 
 $M_R$ to give the Higgs mass of 125 GeV. 
Here we have varied the top quark pole mass in the range of 
 $M_t=173.2 \pm 0.9$ GeV. 
We see that a Higgs mass of 125 GeV is compatible with a seesaw 
scale in the TeV range, in which case the SU(2) triplet neutrinos may be found at the LHC \cite{typeIII-LHC}.

\section{Gauge-Higgs Unification Scenario}
Another scenario for reducing the 
SM vacuum stability bound is to introduce some new physics which effectively lowers the UV cutoff scale below the reduced Planck mass. 
There are several models for achieving this. 
For example, in the Randall-Sundrum model \cite{RS}, 
 the UV cutoff of the SM (as an effective 4-dimensional theory) can be dramatically reduced to $ \Lambda = \omega M_P$ 
 by the `warp factor' $\omega \ll 1$, without too much fine-tuning 
 of the model parameters. 
Alternatively, in the presence of ${\cal N}$ elementary particle species 
 in an effective quantum field theory, 
 the consistency of large-distance black hole physics 
 imposes the following gravitational cutoff on the
 theory, $\Lambda= M_P/\sqrt{\cal N}$ \cite{Dvali}. 
Finally, if we introduce a non-minimal gravitational coupling, 
 $ \xi \Phi^\dagger \Phi {\cal R}$, 
 it seems natural to adopt $\Lambda = M_P/\xi$ 
 \cite{Naturalness} for the effective gravitational cutoff scale. 
Here $\xi$ is a dimensionless coupling constant, $\Phi$ is 
 the SM Higgs doublet, and ${\cal R}$ is the scalar curvature. 
In these scenarios, the SM is realized as an effective theory 
 below the cutoff, so that the RGE of Higgs self-coupling 
 remains the same\footnote{
To be precise, the non-minimal gravitational coupling 
 slightly modifies the SM RGEs, but this effect 
 on the vacuum stability bound is found 
 to be negligible \cite{non-minHOS}.
}. 
However, the cutoff can be considerably below the reduced Planck mass and as a result, the vacuum stability 
 bound on the Higgs mass is reduced. 
The Higgs mass as a function of the effective cutoff 
 $\Lambda$ is depicted in Fig.~2. 
A Higgs mass $m_H=125$ GeV can be realized with an effective cutoff $\Lambda \simeq 1.4 \times 10^{11}$ GeV, 
 which corresponds to $\omega^{-1}$, $\sqrt{\cal N}$, ${\xi} \simeq 10^7$.

\begin{figure}[t]
\begin{center}
{\includegraphics*[width=.7\linewidth]{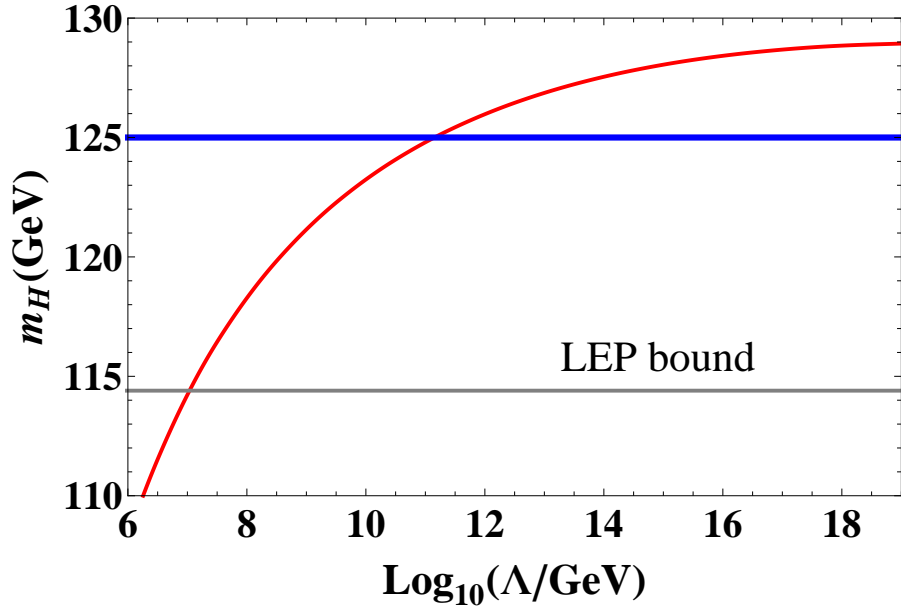}}
\caption{
The vacuum stability bound on the Higgs mass
 versus the effective cutoff scale $\Lambda$, for $M_t =173.2$ GeV. 
A Higgs mass of 125 GeV corresponds to $\Lambda\simeq1.4 \times 10^{11}$ GeV. 
}
\end{center}
\end{figure}

The vacuum stability bound is the lower bound 
 on the Higgs boson mass obtained with a fixed cutoff scale. 
Thus, Fig.~2 indicates that to achieve $m_H=125$ GeV, the upper bound on the effective UV cutoff is $\Lambda \lesssim 1.4 \times 10^{11}$ GeV.  
It is therefore interesting to see if there exists models 
 which can predict the Higgs boson mass once the effective UV cutoff is fixed. 
The gauge-Higgs unification scenario \cite{GHU1, GHU2} 
 provides a good example of one class of such models. 

In the gauge-Higgs unification scenario, the SM Higgs doublet 
 is identified as the extra-dimensional component of  
 a higher dimensional gauge field, so that 
 the Higgs self-coupling is determined by the gauge invariance 
 in higher dimensions. 
As has been shown in Ref.~\cite{HMOY}, 
 the low energy effective theory of the gauge-Higgs unification scenario 
 is equivalent to the SM with a certain boundary condition 
 for the Higgs self-coupling imposed at the compactification scale 
 of the extra-dimensions, the so-called ``gauge-Higgs condition''. 
In particular, in the gauge-Higgs unification scenario in flat 5D spacetime, the gauge-Higgs condition requires a vanishing Higgs self-coupling 
 at the cutoff scale $\Lambda$, which is identified as the compactification scale. 
Therefore, in a gauge-Higgs unification scenario, the vacuum stability bound 
 on the SM Higgs boson mass is simply the Higgs mass prediction 
 with the compactification scale $\Lambda$. 
Based on this identification, the SM Higgs boson mass has been
 calculated some time ago, in Ref. \cite{GHU-Hmass} .

\begin{figure}[t]
\begin{center}
{\includegraphics*[width=.7\linewidth]{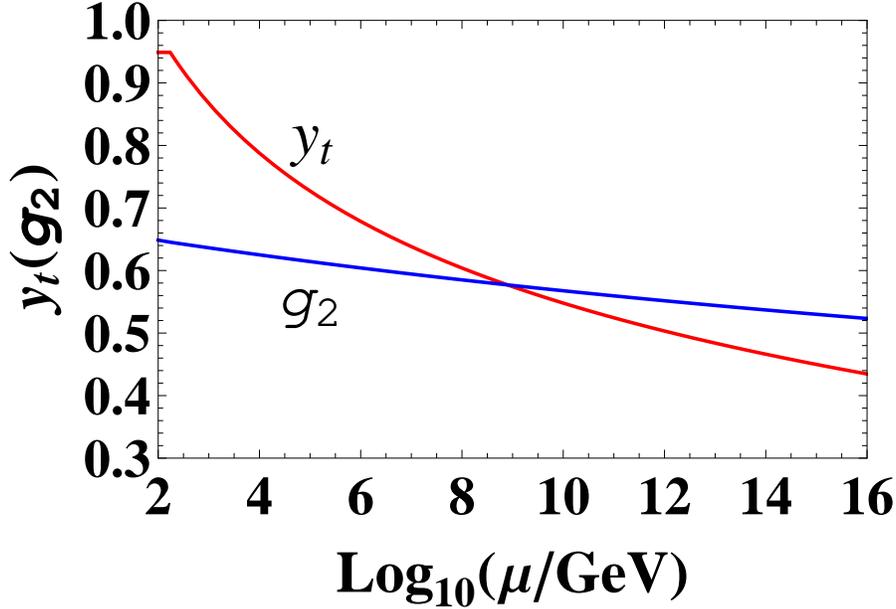}}
\caption{
RGE running of the SU(2) gauge coupling and top Yukawa coupling. 
They unify at $7.9 \times 10^8$ GeV, 
 which is identified as the compactification scale. 
}
\label{g2yt}
\end{center}
\end{figure}

In the following, inspired by the 125 GeV Higgs, we re-calculate the Higgs mass prediction 
 in the 5D gauge-Higgs scenario. 
We have improved upon the previous analysis in \cite{GHU-Hmass} 
 by taking into account the two-loop RGE improved 
 one-loop effective Higgs potential in Eq.~(\ref{HiggsPole}) 
 and the updated top quark pole mass $M_t=173.2 \pm 0.9$ GeV. 
From the viewpoint of the gauge-Higgs unification scenario, 
 the result shown in Fig.~2 indicates that 
 a compactification scale of $\Lambda\simeq1.4 \times 10^{11}$ GeV 
 results in $m_H=125$ GeV.

Since the gauge Higgs unification scenario also provides 
 unification of the gauge coupling and top quark Yukawa coupling 
 at the compactification scale, 
 we may identify the compactification scale 
 with the unification scale of the SU(2) gauge coupling ($g_2$) 
 and top Yukawa coupling ($y_t$) (see Fig.~\ref{g2yt}). 
In this way we predict the Higgs boson mass, as shown in Table~2 for varying top quark mass 
 and the compactification scale.

\begin{table}[ht]
\begin{center}
\begin{tabular}{|c|c|c|}
\hline
$M_t$ (GeV) & $g_2-y_t$ unification scale (GeV) & Predicted $m_H$ (GeV) \\ 
\hline
$172.3$ & $4.3 \times 10^8$ & 118.6  \\ 
\hline
$173.2$ & $7.9 \times 10^8$ & 120.9  \\ 
\hline
$174.1$ & $1.5 \times 10^9$ & 123.2  \\
\hline
\end{tabular}
\end{center}
\caption{
Predicted Higgs boson mass for varying top quark pole mass, 
 with the compactification scale determined by 
 the unification of the SU(2) gauge coupling and top Yukawa coupling. 
}
\end{table}

\section{Conclusion}

In conclusion, an excess of events around $125$ GeV 
 recently reported by the ATLAS and CMS experiments may be the first evidence for the SM Higgs boson. 
We have considered possible implications of a 125 GeV Higgs 
 for new physics from the viewpoint of the vacuum stability  
 bound on the SM Higgs mass with the reduced Planck mass cutoff. 
Since the (absolute) vacuum stability bound is close to 129 GeV, 
 some new physics is needed to bring it down to 125 GeV. 
We first considered the seesaw extension of the SM 
which incorporates the observed neutrino masses and oscillations. 
In this case, the RGE of the SM Higgs self-coupling is modified for energies higher than the seesaw scale. 
In type II and type III seesaw, the 125 GeV mass can be achieved 
 with the seesaw scale much lower than the conventional 
 intermediate scale. 
With type III seesaw, the vacuum stability bound on the Higgs mass  
 can be lowered to 125 GeV with 2 or 3 generations of SU(2) triplet neutrinos, with the seesaw scale as low as a TeV.

If there is no modification of the SM RGEs, 
 it is necessary to introduce an effective ultraviolet cutoff $\Lambda \simeq 10^{11}$ GeV to
 yield a vacuum stability bound of $m_H = 125$ GeV. 
We have discussed various new physics models which provide 
 such a low cutoff scale. 
In a gauge-Higgs unification 
 scenario in 5D flat spacetime, the vacuum stability 
 bound of $m_H = 125$ GeV is identified as a prediction 
 of the Higgs mass under the gauge-Higgs condition imposed at 
 the compactification scale $\Lambda \simeq 10^{11}$ GeV. 
If the compactification scale is identified with 
 the unification scale of the SU(2) gauge coupling
 and top Yukawa coupling, the Higgs mass is predicted to lie
  close to 125 GeV.

Finally, while we have required absolute electroweak vacuum stability 
 in this paper, one may loosen the bound by considering 
 meta-stability, which leads to a lower bound $m_H \gtrsim 110$ GeV on the Higgs mass.
[See \cite{Meta-STB} for recent analysis in this context.]
In this case, we may consider new physics effects 
 which raise the bound to 125 GeV. 
As has been shown in \cite{HMass-typeI, HMass-typeIII}, 
 type I and type III seesaw with a seesaw scale $\gtrsim 10^{14}$ GeV 
will do this.

\section*{Acknowledgments}
This work is supported in part by the DOE Grants, 
 \# DE-FG02-12ER41808 (B.H. and Q.S.) 
 and \# DE-FG02-10ER41714 (N.O.), 
 and by Bartol Research Institute (B.H.).


\end{document}